\RequirePackage{fix-cm}
\documentclass[smallextended]{svjour3}       
\smartqed  
\usepackage{graphicx}
\begin{document}

\title{Real-space renormalization of randomly vacated lattices: a renormalization group for jamming?
}


\author{Abram H. Clark   
}


\institute{A. Clark \at
              Naval Postgraduate School \\
              Department of Physics \\
              Monterey, CA 93943 \\
              \email{abe.clark@nps.edu}           
}

\date{Received: date / Accepted: date}

\maketitle

\begin{abstract}
Jamming occurs in granular materials, as well as in emulsions, dense suspensions, and other amorphous, particulate systems. When the packing fraction $\phi$, defined as the ratio of particle volume to system volume, is increased past a critical value $\phi_c$, a liquid-solid phase transition occurs, and grains are no longer able to rearrange. Previous studies have shown evidence of spatial correlations that diverge near $\phi = \phi_c$, but there has been no explicit spatial renormalization group (RG) scheme that has captured this transition. Here, I present a candidate for such a scheme, using a block-spin-like transformation of a randomly vacated lattice of grains. I define a real-space RG transformation based on local mechanical stability. This model displays a critical packing fraction $\phi_c$ and gives estimates of critical exponents in two and three dimensions.
\end{abstract}

\section*{Introduction}
\label{intro}

Jamming~\cite{jaeger1996,ohern2003,donev2004} describes a liquid-solid phase transition that occurs as amorphous, athermal systems are compressed from a dilute state. Jamming is often used to analyze dense granular materials~\cite{majmudar2007jamming,dauchot2005dynamical,behringer2014}, as well as foams~\cite{durian1995foam}, emulsions~\cite{paredes2013rheology}, and colloids~\cite{Nordstrom2010}. Additionally, jamming is used to study glassy dynamics more broadly~\cite{silbert2002analogies}. The canonical model system for jamming is an ensemble of soft, frictionless spheres, where the packing fraction $\phi$, defined as the ratio of total volume of the particles to total system volume, is varied. Disordered solid-like states can form at and above a packing fraction $\phi_c$, which is smaller than the density of a close-packed crystal, $\phi_{\rm xtal}$. In three dimensions, $\phi_{\rm xtal} = \pi / 3\sqrt{2} \approx 0.74$ and $\phi_c \approx 0.64$. Several studies~\cite{ohern2003,olsson2007,Nordstrom2010,olsson2011,vaagberg2011finite,paredes2013rheology,goodrich2016scaling} have shown evidence for a diverging length scale, $\xi \propto |\phi - \phi_c|^{-\nu}$, that controls the mechanical response near $\phi \approx \phi_c$, suggesting that jamming is a kind of non-equilibrium critical transition. 

There is currently no real-space renormalization group (RG) scheme~\cite{kadanoff1966scaling,wilson1971renormalization,wilson1971renormalization2} that captures this diverging length scale. The disordered nature of solid-like states near $\phi_c$ makes a block-spin approach~\cite{kadanoff1966scaling} impossible, since there is no lattice on which to define a block-spin-like coarsening and rescaling. One possible way around this problem is to consider a specially prepared state of a randomly vacated lattice (RVL) of repulsive spheres, where each site is occupied with some probability. If this system is then compressed or thermalized, the spheres may rearrange locally, based on whether their neighbors are occupied. These local mechanical instabilities can couple together spatially and lead to global relaxation to a liquid-like state. This should occur at some $\phi=\phi_c$ such that $0<\phi_c<\phi_{\rm xtal}$. It is not obvious how this $\phi_c$, defined as the packing fraction of the RVL state that will fully relax to a liquid, relates to $\phi_c$ for typical jamming studies, which are formed by, e.g., quasistatic compression of a dilute system of particles with random initial positions. I provide a more substantial discussion of the possible connection between RVL states and jamming of disordered sphere packings below at the end of the paper.

If $\phi_c$ for the RVL system does in fact capture the same critical point as the jamming packing fraction, then the lattice symmetry of the RVL system would allow a block-spin-like RG transformation, which is the focus on this paper. Here, I propose such an RG transformation and solve it in both two (2D) and three (3D) dimensions. Blocks of lattice sites are coarse-grained into a super-lattice, and each super-site is either ``occupied'' or ``unoccupied'' based on mechanical stability of the underlying spheres that make it up. This approach is similar to a class of hierarchically (or kinetically) constrained models~\cite{Palmer1984,Fredrickson1984,Ritort2003,Whitelam2005}, but here the RG transformation is based on local geometric stability. If the super-sites are assumed to interact with each other in the same way as the original sites, then this process is repeatable. This model yields a value for $\phi=\phi_c$, corresponding to the packing fraction at which the lattice appears statistically unchanged after an RG transformation, as well as a critical exponent $\nu$. The values of $\phi_c$ and $\nu$ are similar to the accepted values for jamming in 2D and 3D, but future work is needed to fully establish the relevance of these results.

\section*{The RVL model and RG}

\begin{figure}
\raggedright
(a) \hspace{37mm} (b) \hspace{37mm} (c) \\
 
\includegraphics[width=0.45\textwidth]{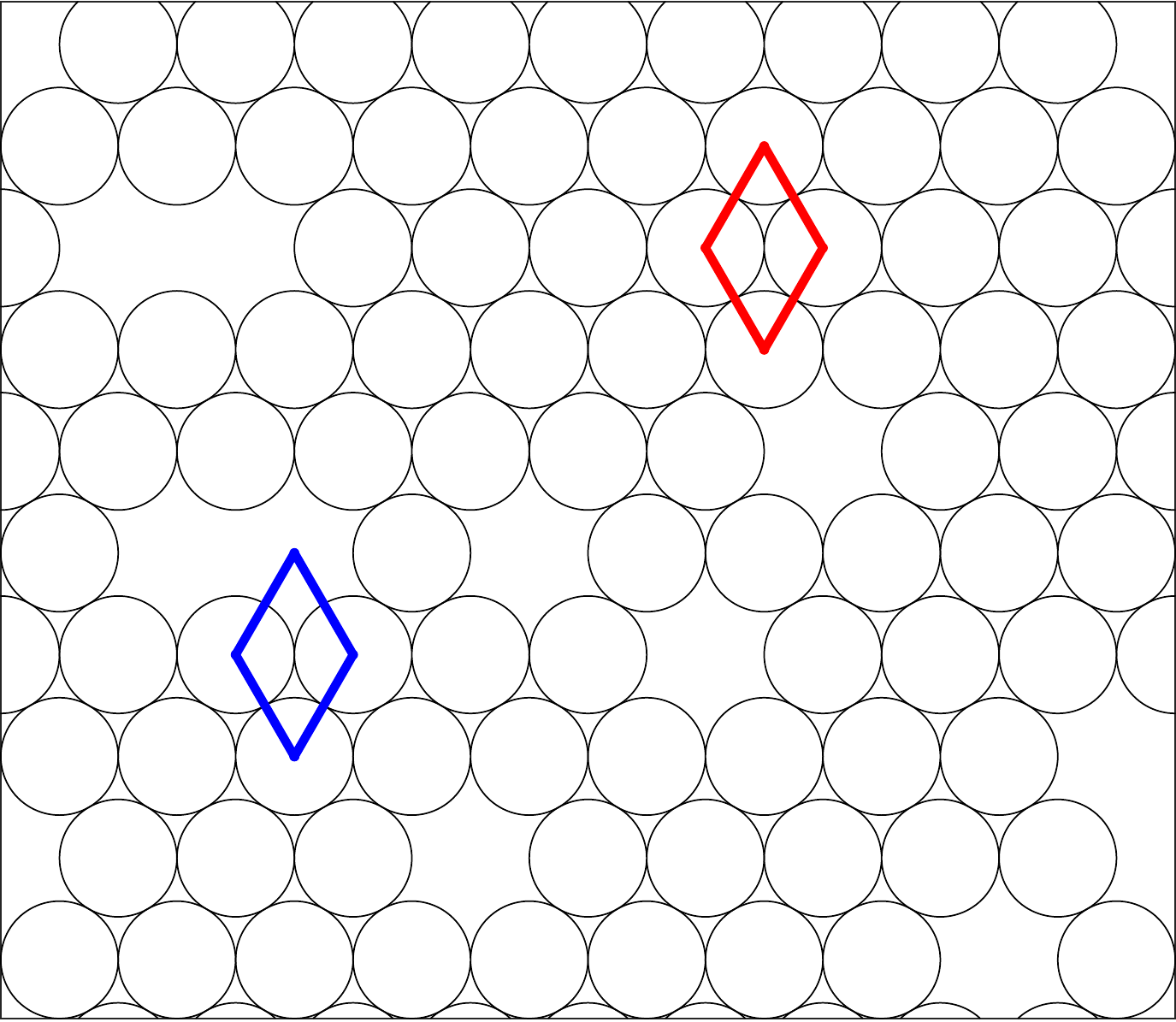}
\includegraphics[width=0.45\textwidth]{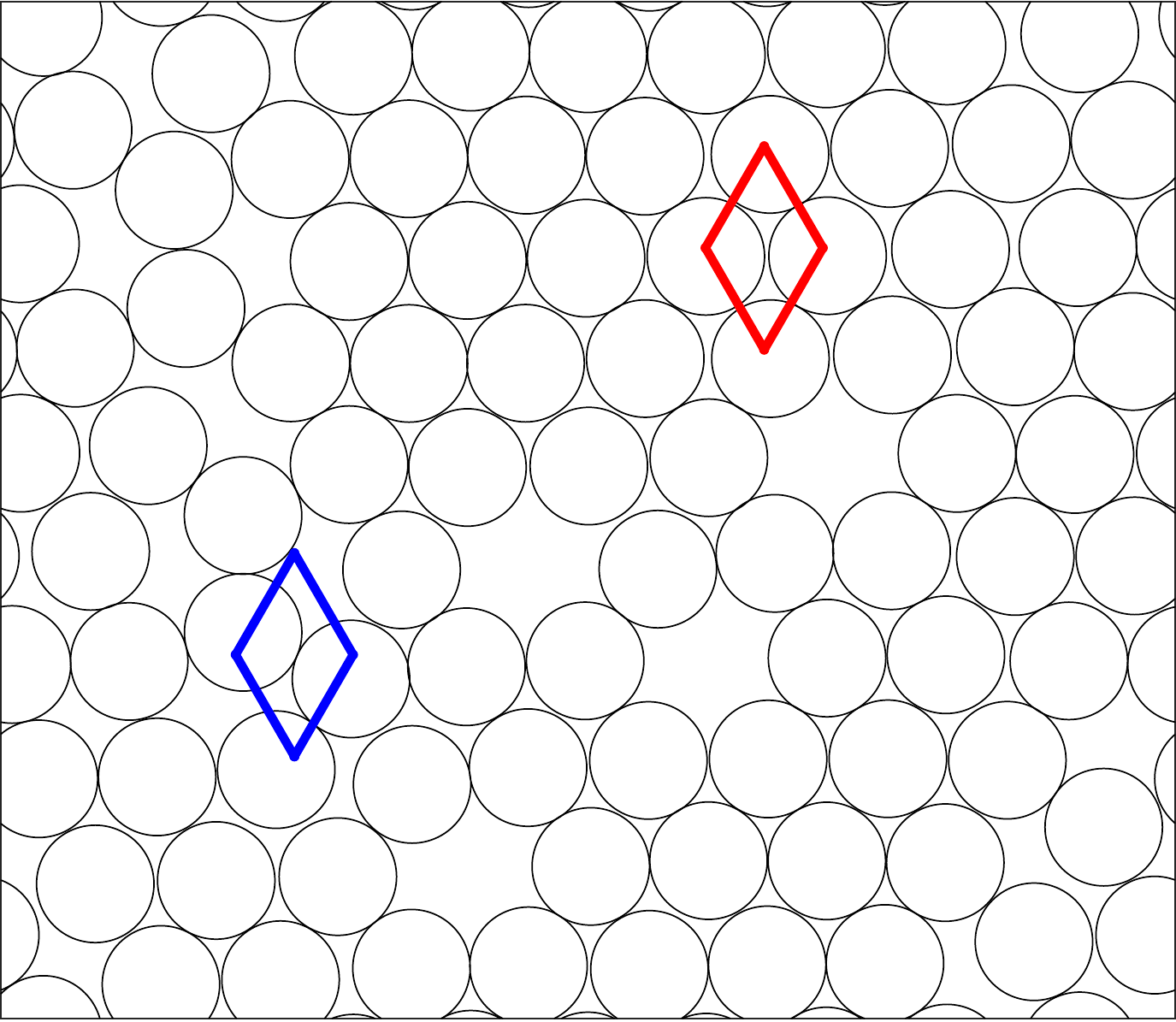}
\caption{(a) A hexagonal lattice in 2D with 88 occupied sites and 12 empty sites, corresponding to $p=0.88$. There are three pairs of empty neighboring sites (unstable), and six single unoccupied sites (stable). The diamonds show two super-sites as defined in the text, one stable (red) and one unstable (blue). (b) The same system, but evolved using molecular dynamics. Note that the single unoccupied sites are stable, except for the ones near a pair of unoccupied sites. (c) The unit cell for the real-space RG transformation solved in 2D, where red circles, labeled A and B, comprise the core unit cell with neighboring blue circles labeled 1, 2, and 3.}
\label{fig: MD}
\end{figure}

An RVL state consists of repulsive spheres (or disks) arranged on a close-packed lattice with some fraction of the spheres removed randomly, such that a single lattice site is occupied with probability $p = \phi/\phi_{\rm xtal}$. Local mechanical instability occurs when multiple neighboring sites are unoccupied~\cite{torquato2007toward,stillinger2003,Kansal2002}. For example, a single missing particle in the 2D hexagonal lattice (see Fig.~\ref{fig: MD}) is still stable (solid-like), since the six neighbors block each other from entering the unoccupied site. If two neighboring sites are unoccupied, then the nearby region will be liquid-like, and particles on either side of this missing pair can move into the void~\cite{stillinger2003}. In the 3D face-centered cubic (FCC) lattice, local instability requires a ``missing triad'' of unoccupied sites, with three unoccupied neighboring sites forming an equilateral triangle~\cite{torquato2007toward}. 

Each local instability will affect its neighbors over some length scale, and these instabilities can couple together over many length scales to yield global liquid-like behavior~\cite{Palmer1984}, suggesting the need for an RG approach. The technique of a real-space RG transformation, pioneered by Kadanoff~\cite{kadanoff1966scaling} and later fully developed by Wilson~\cite{wilson1971renormalization,wilson1971renormalization2}, consists of two essential steps. First, the system is coarsened over some length scale $b$, and degrees of freedom on length scales smaller than $b$ are averaged out. Second, the resulting system is spatially rescaled (i.e., zoomed out) by a factor $b$ such that the rescaled system has the same microscopic length scale as the original system. For jamming, these two steps would be represented by a transformation $\phi' = R_b(\phi)$, where $R_b$ relates the effective packing fraction $\phi'$ of the system after the RG transformation to the packing fraction $\phi$ of the original system. At the critical packing fraction $\phi_c$, the system is correlated over infinitely large distances. Thus, at $\phi=\phi_c$ the system should be statistically invariant under an RG transformation, i.e., $\phi_c = R_b (\phi_c)$.

In the block-spin system, Kadanoff assumed that the coarse-grained Hamiltonian has the same form as the original Hamiltonian and that super-sites interact with each other in the same way as the underlying spins. In reality, there is a more complicated rule that relates the coarse-grained Hamiltonian and interactions to the original, but Kadanoff's approximation showed how, if such a rule was known, the full RG picture could be formalized. Here, I make similar assumptions: I define a block-spin-like RG transformation by grouping the lattice into super-sites consisting of $b^D$ individual sites, where $D$ is the dimensionality. Each super-site is mechanically stable or unstable depending on the criteria outlined above. After coarse-graining, mechanically stable super-sites are denoted as occupied, and the procedure can be repeated at the next level. Thus, super-sites are assumed to interact in the same way as individual particles. Ideally, the super-sites should fill space in the same way as the initial lattice, since the procedure should be repeatable. 

Using these assumptions, I define a rule $R_b$ relating the occupation probability $p'$ of the super-site to $p$:
\begin{equation}
p'=R_b(p)
\label{eqn:RGxform}
\end{equation} 
The critical occupation probability $p_c$ corresponds to the non-trivial state (i.e. not completely occupied or unoccupied) at which the RG transformation leaves the system statistically unchanged.
\begin{equation}
p_c=R_b(p_c)
\label{eqn:RGcrit}
\end{equation}
Assuming that there is a diverging length scale $\xi \propto |\phi-\phi_c|^{-\nu} \propto |p-p_c|^{-\nu}$ and that $R_b$ reduces the correlation length $\xi$ by a factor of $b$, i.e., $\xi' = \frac{\xi}{b} \propto |p'-p_c|^{-\nu}$, one can write:
\begin{equation}
| R_b(p)-p_c | ^{-\nu} = \frac{| p - p_c | ^{-\nu}}{b}.
\end{equation}
Rearrangement combined with Eq.~(\ref{eqn:RGcrit}) yields an expression for $\nu$:
\begin{equation}
\nu = \frac{\log b}{\log \frac{| R_b(p)-p_c|}{| p-p_c|}}=\frac{\log b}{\log \frac{| R_b(p)-R_b(p_c)|}{| p-p_c|}},
\label{eqn:nu1}
\end{equation}
In the limit of $p\rightarrow p_c$, this becomes:
\begin{equation}
\nu = \frac{\log b}{\log \frac{dR_b(p)}{dp}| _{p=p_c}}.
\label{eqn:nu2}
\end{equation}
So, any choice of $R_b$ yields a value of the critical point, $p_c$, and the value of the critical exponent, $\nu$.

\section*{Solution in two dimensions}
\begin{figure}
    \includegraphics[width=0.8\textwidth]{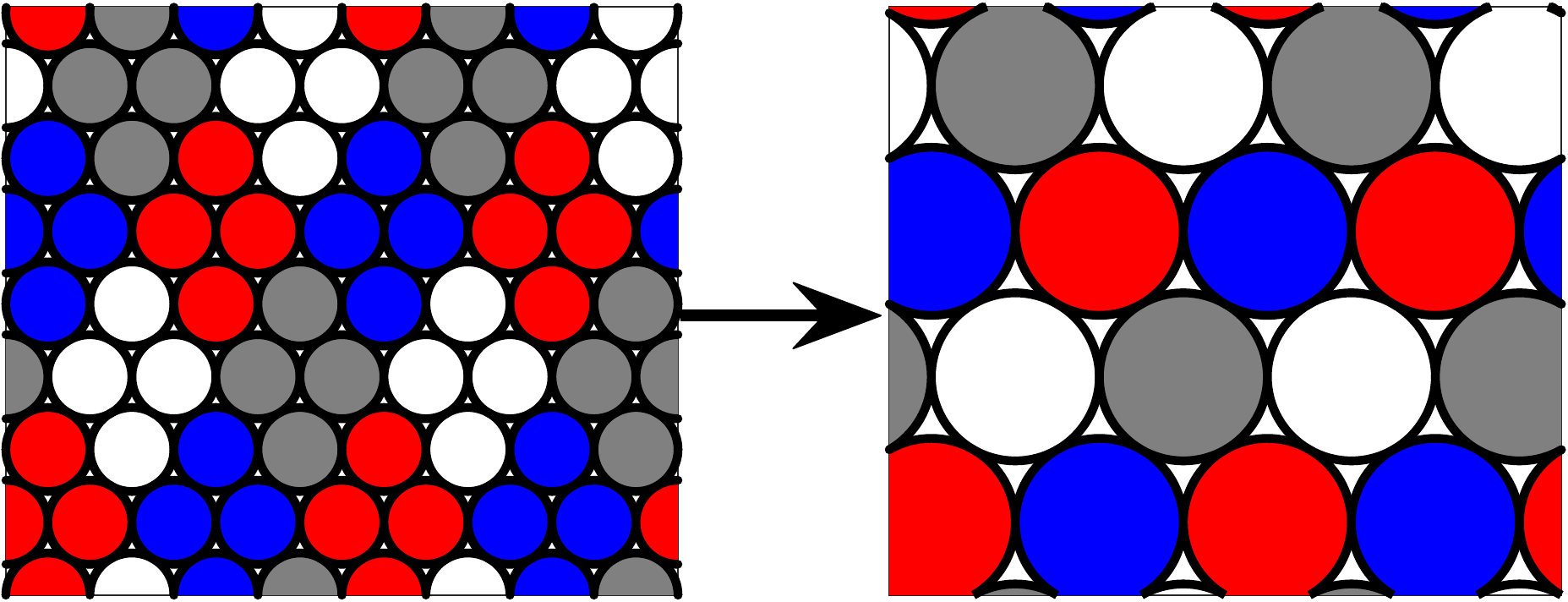}
    \caption{A graphical representation of the coarse-graining procedure in the RG transformation. Each colored four-site diamond on the left becomes a single super-site (with the same color) on the right.}
    \label{fig:hexlatticecolors2}
\end{figure}

For the 2D lattice, I choose to define super-sites as four-site diamond shapes, shown in Fig. \ref{fig:hexlatticecolors2}. As I discuss below, this choice is not unique and does affect the results of the following calculation. Upon application of the RG transformation, each four-site cluster would be represented by a single super-site which would be stable (occupied) or unstable (unoccupied). Since instability requires two neighboring sites to be unoccupied, stability for a cluster can then be determined by simply looking at possible configurations and applying this rule. Figure~\ref{fig: MD}(a) shows an RVL system in 2D, and Fig. \ref{fig: MD}(b) shows the same system evolved with molecular dynamics for some time. Note that regions with missing pairs have become liquid-like, but single missing sites remain solid-like, unless they are near a missing pair. Since $b=2$ for the four-site diamond cell, I denote this RG transformation as $R_{b=2}^{\rm (2D)}$.

Before writing $R_{b=2}^{\rm (2D)}$ and solving for the fixed point, I first note a few properties of this particular choice of a unit cell. (1): The four sites that make up a super-site are not not rotationally symmetric. Thus, for single super-sites, the RG transformation introduces a kind of strain, where the vertical direction is contracted relative to the horizontal direction, converting an asymmetric diamond-shape into a symmetric circular shape. (2): Despite (1), the resulting super-sites perfectly tile the lattice, which is apparent from Fig.~\ref{fig:hexlatticecolors2}. (3): Additionally, the lattice symmetry of the super-sites is identical to the original lattice symmetry (i.e., the RG transformation retains the same Bravais vectors as the original lattice). This means that the vertical contraction in (1) does not apply to the lattice as a whole. This can be seen in Fig.~\ref{fig:hexlatticecolors2} by the fact that the centers of the supersites form an equilateral triangle both before and after the coarse-graining is applied. Properties (2) and (3) are philsosophically appealing, since an RG transformation should be repeatable, and thus the lattice of super-sites should be identical to the original lattice.

\begin{figure}
    \includegraphics[width=0.4\textwidth]{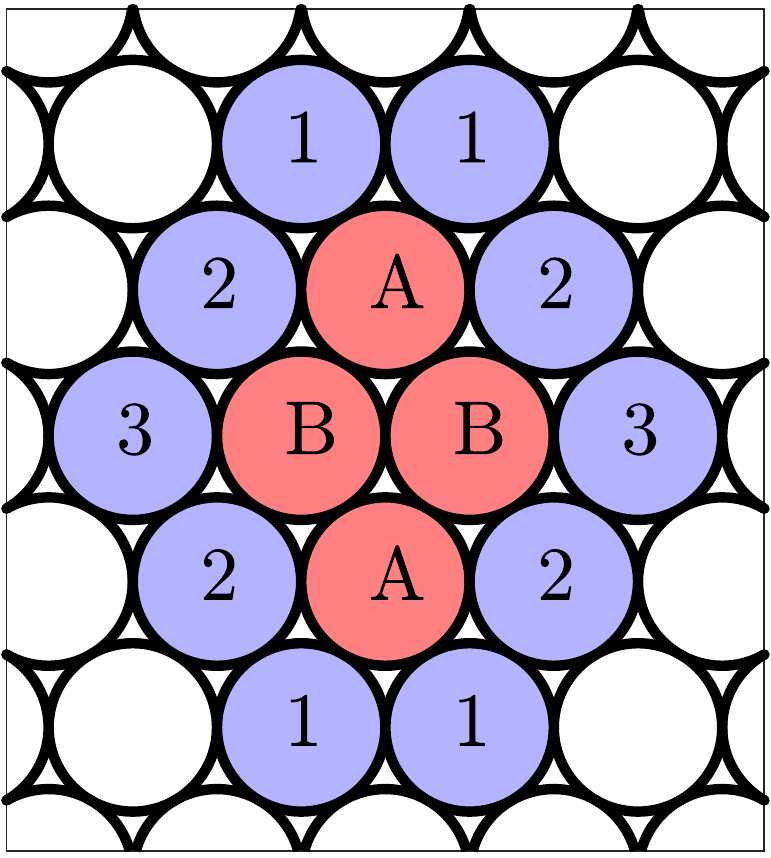}
    \caption{The four-site unit cell is indicated by the red circles labeled A or B. The neighboring sites, which are used in the calculation leading to Eq.~(\ref{eqn:2Drenorm}), are indicated by blue circles labeled 1, 2, or 3.}
    \label{fig:unitcell}
\end{figure}

To explicitly write $R_{b=2}^{\rm (2D)}$ for the four-site diamond-shaped cluster, I consider a four core sites and all nearest neighbors (i.e. particles which could move into the cluster's lattice sites), as shown in Fig. \ref{fig:unitcell}. The central cluster particles are labeled either A or B, and the neighboring particles are labeled 1-3 (here, referred to as N1, N2, and N3). Also, note that $b=2$ for this configuration, since our two-dimensional super-site contains four particles. If all four sites are occupied, which occurs with probability $p^4$, then the super-site is stable. If only three sites are occupied, then these configurations can either be stable or unstable depending on whether the nearest-neighbor sites are occupied. A single unoccupied A-site occurs with probability $2p^3(1-p)$, and stability requires all four associated N1 and N2 sites to be occupied, which occurs with probability $p^4$. A single unoccupied B-site occurs with probability $2p^3(1-p)$, and stability requires the N3 and both N2 sites to be occupied, which occurs with probability $p^3$. Next, the configuration with two occupied B-sites and two unoccupied A-sites occurs with probability $p^2(1-p)^2$. Stability for this configuration requires all N1 and N2 sites to be occupied, which occurs with probability $p^8$. No other configurations can be stable. 
The total probability of stability is thus:
\begin{equation}
p'=R_{b=2}^{\rm (2D)}=p^4+2p^3(1-p)(p^4+p^3)+p^2(1-p)^2p^8
\label{eqn:2Drenorm}
\end{equation}
\begin{figure}
    \includegraphics[scale=0.5]{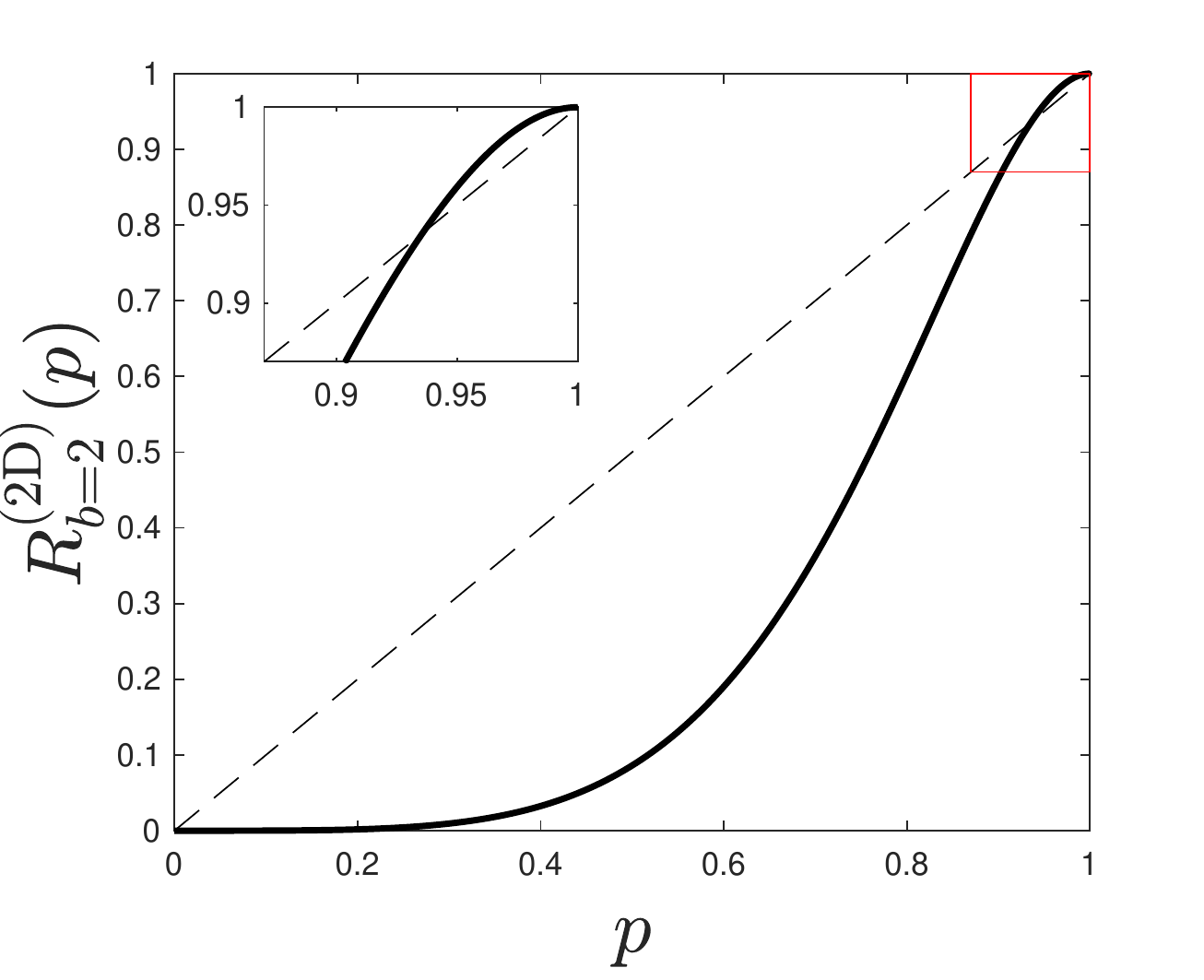}
    \caption{The function $R_{b=2}^{\rm (2D)}$ given in Eq.~(\ref{eqn:2Drenorm}) is plotted with the solid black line. The dashed line shows a line of slope 1, meaning that two curves cross at $p=p_c$. The region outlined in red is shown in the inset, and the crossing point is consistent with Eq.~(\ref{eqn:pc}).}
    \label{fig:R_b_2D}
\end{figure}
This function is plotted in Fig.~\ref{fig:R_b_2D} along with a dashed line representing $p'=p$. These lines cross three places: at $p=0$ and $p=1$, representing the trivial fixed points (totally unoccupied or totally occupied), and the nontrivial critical point where $R_{b=2}^{\rm (2D)}(p_c)=p_c$. Solving numerically yields
\begin{equation}
p_c \approx 0.9356.
\label{eqn:pc}
\end{equation}
Thus, the critical packing fraction is given by
\begin{equation}
\phi_{c,2D} =p_c\phi_L \approx 0.8485,
\label{eqn:phic2D}
\end{equation}
where $\phi_L=\pi\sqrt{3}/6$ is the packing density of a fully occupied lattice. This value is similar to the packing fraction $\phi_{J,2D}\approx 0.84$~\cite{ohern2003,olsson2007} where jamming occurs for disordered packings of frictionless disks in 2D. Equation \ref{eqn:nu2} yields a value for the critical exponent, $\nu$.
\begin{equation}
\nu = \frac{\log b}{\log \frac{dR_b(p)}{dp}\vert _{p=p_c}}=\frac{\log 2}{\log 1.7942...}=1.1857...
\label{eqn:nu2D}
\end{equation}
Compare this value to the exponent obtained in 2D by V{\aa}gberg, et al.~\cite{vaagberg2011finite}, $\nu\approx 1$. This value was obtained by considering corrections to scaling (i.e., by considering irrelevant variables that only matter for small system sizes). Other studies have also found $\nu$ between 0.6 and 0.7~\cite{ohern2003,olsson2007} in 2D when corrections to scaling were not considered. Additionally, other estimates of $\nu$ exist for other quanitites, e.g., Refs.~\cite{wyart2005geometric,silbert2005}, and a full RG description should be able to relate these different divergences.

As mentioned above, the four-site diamond is certainly not the only choice of unit cell. For example, a nine-site diamond would also tile the unit cell in the same way. Also, a three-site triangle and a seven-site hexagon super-site will both tile the lattice, but they require a spatial rotation after coarse-graining to make the super-lattice identical to the original lattice. In fact, the combinatorics described above can be applied to any cluster of sites that does not (in general) tile the lattice or retain the lattice symmetry after coarse-graining. The critical packing fraction $\phi_c$ and diverging length scale exponent $\nu$ vary somewhat with the choice of unit cell. For example, a three-site triangular cluster has $b=\sqrt{3}$ and $p' = p^3 + 3p^2(1-p)p^4$ (the first term corresponds to three occupied sites, and the second term corresponds to one missing site with three-fold multiplicity, which requires the four neighbors of that site to be occupied). This yields $p_c \approx 0.914$, $\phi_c \approx 0.829$, and $\nu \approx 0.988$. Similarly, the seven-site hexagon, with $b=\sqrt{7}$, has $p' = p^7+p^6(1-p)(1+6p^3)+p^5(1-p)^2(9p^6)+p^4(1-p)^3 (2p^9)$, yielding $\phi_c \approx 0.8724$ and $\nu \approx 1.565$. So, increasing $b$ tends to increase $\phi_c$, which must be the case since stability will get increasingly difficult for bigger clusters. In the limit of large unit cell, $p_c = 1$, since any $p<1$ will lead to a local instability somewhere in an infinite system.

\section*{Solution in three dimensions}

\begin{figure}
\raggedright \hspace{2mm} (a) \hspace{45mm} (b) \hspace{20mm} (c) \\   \includegraphics[width=\textwidth]{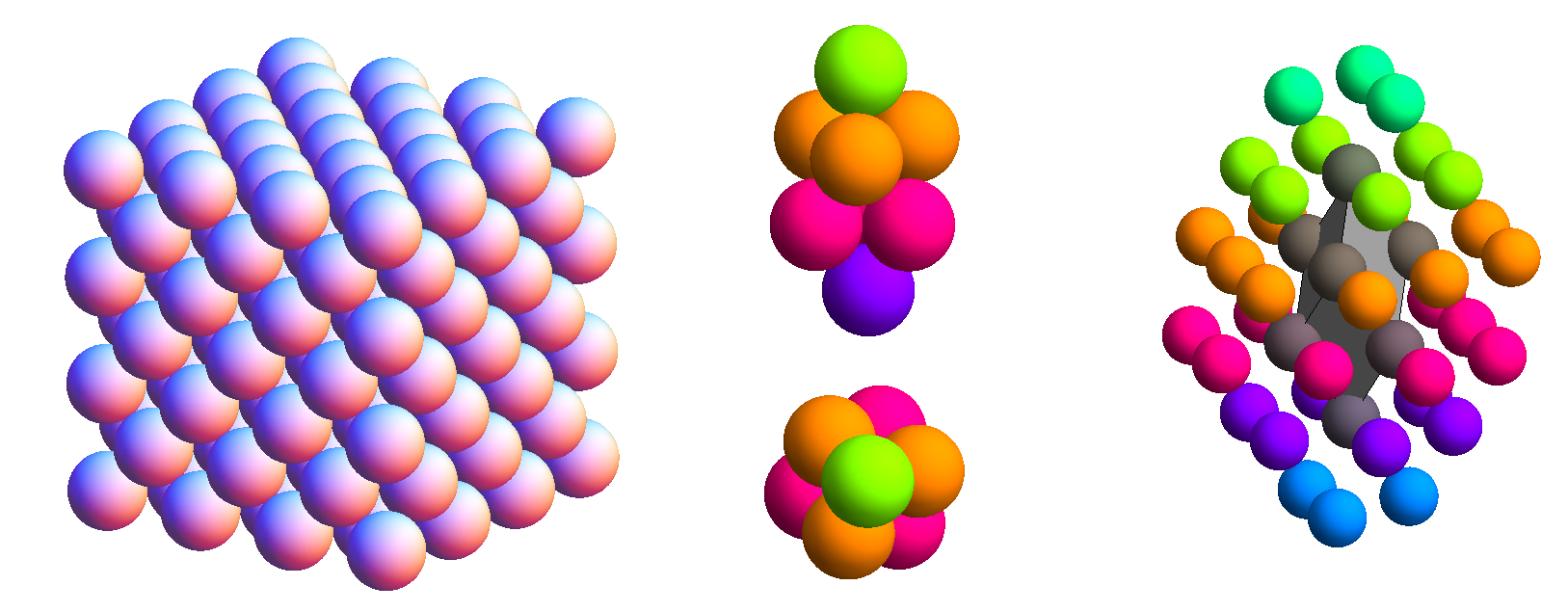}
\caption{(a), An FCC lattice of close-packed spheres, where the edge-faces are showing the square layers, not the hexagonal layers. (b) The eight-particle core unit for renormalization (analogous to the four-particle diamond in 2D) shown from a side and top view. (c) The eight-particle core unit (in gray) with all 36 neighbors.}
\label{fig:fccunitcell}
\end{figure}

In 3D, there are two ways to pack spheres with the highest possible density, $\phi_{\rm xtal} \approx 0.74$: face-centered cubic (FCC) and hexagonally close-packed (HCP). Both lattices consist of layers of hexagonally packed spheres stacked on top of one another. Here, I consider the FCC lattice, shown in Fig.~\ref{fig:fccunitcell}(a). I group the lattice sites into super-sites as shown in Fig.~\ref{fig:fccunitcell}(b). The super-sites have $2^3=8$ sites, so $b=2$. The eight-cell super-site has 36 neighbors, as shown in Fig.~\ref{fig:fccunitcell}(c). This super-site does tile the lattice with the same symmetry as the lattice itself, but each super-site is rotated 180-degrees with respect to its nearest neighbors.

Local instability requires a missing triad, as discussed above. If a state has a missing triad among the eight core sites, it is unstable. Otherwise, all the neighbors of unoccupied sites must be considered. Every stable configuration occurs with probability $p^n (1-p)^{(44-n)}$, where $n$ is the number of occupied sites in that particular configuration (and 44 is the total number of sites).

\begin{figure}
\raggedright \hspace{2mm} (a) \hspace{18mm} (b) \hspace{18mm} (c) \hspace{18mm} (d) \hspace{18mm} (e) \\ 
\includegraphics[trim=70mm 70mm 70mm 70mm, clip, width=0.19\textwidth]{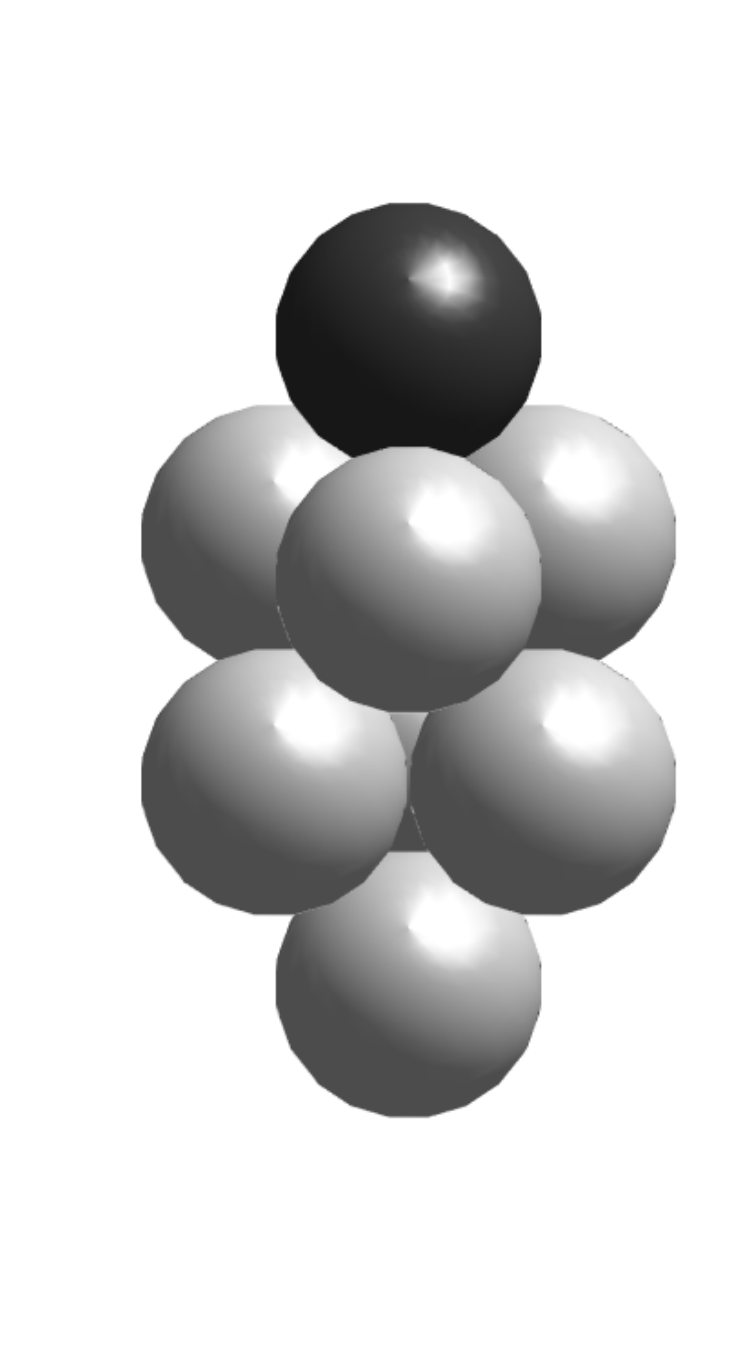}
\includegraphics[trim=70mm 70mm 70mm 70mm, clip, width=0.19\textwidth]{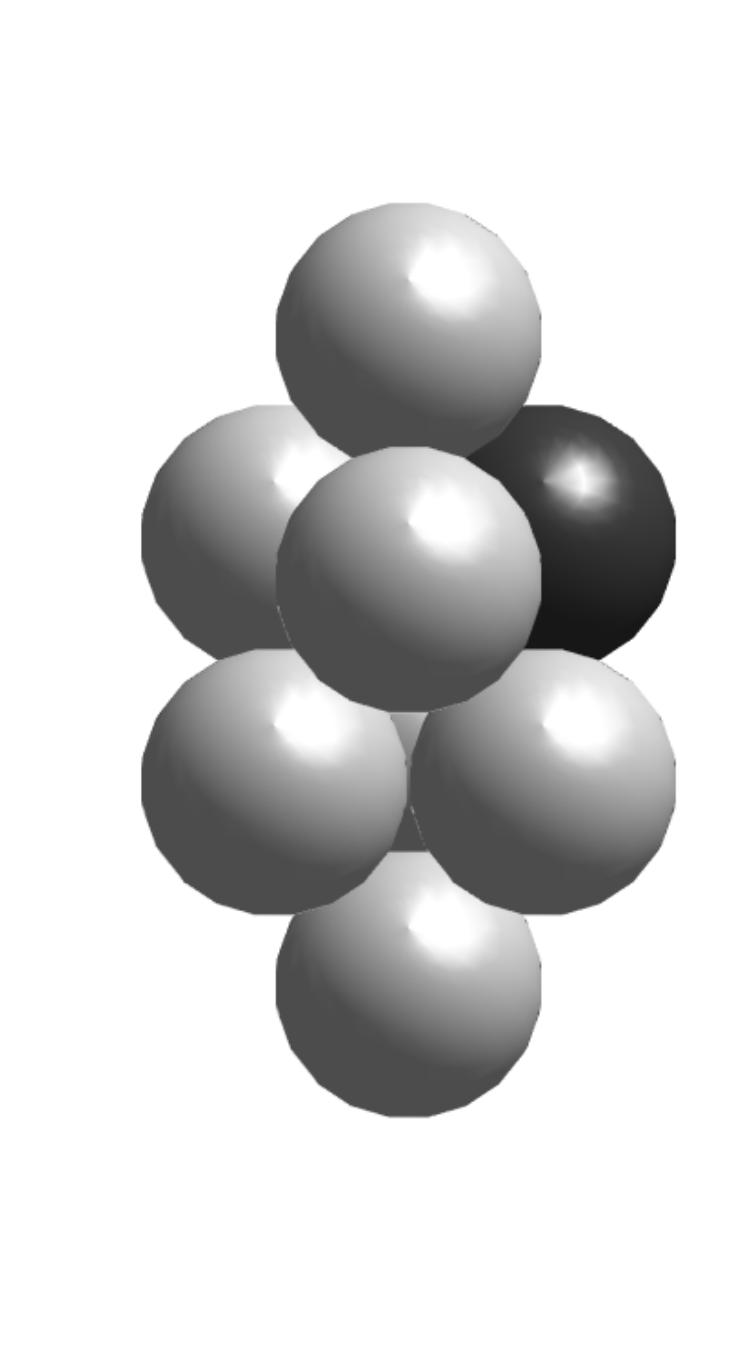}
\includegraphics[trim=70mm 70mm 70mm 70mm, clip, width=0.19\textwidth]{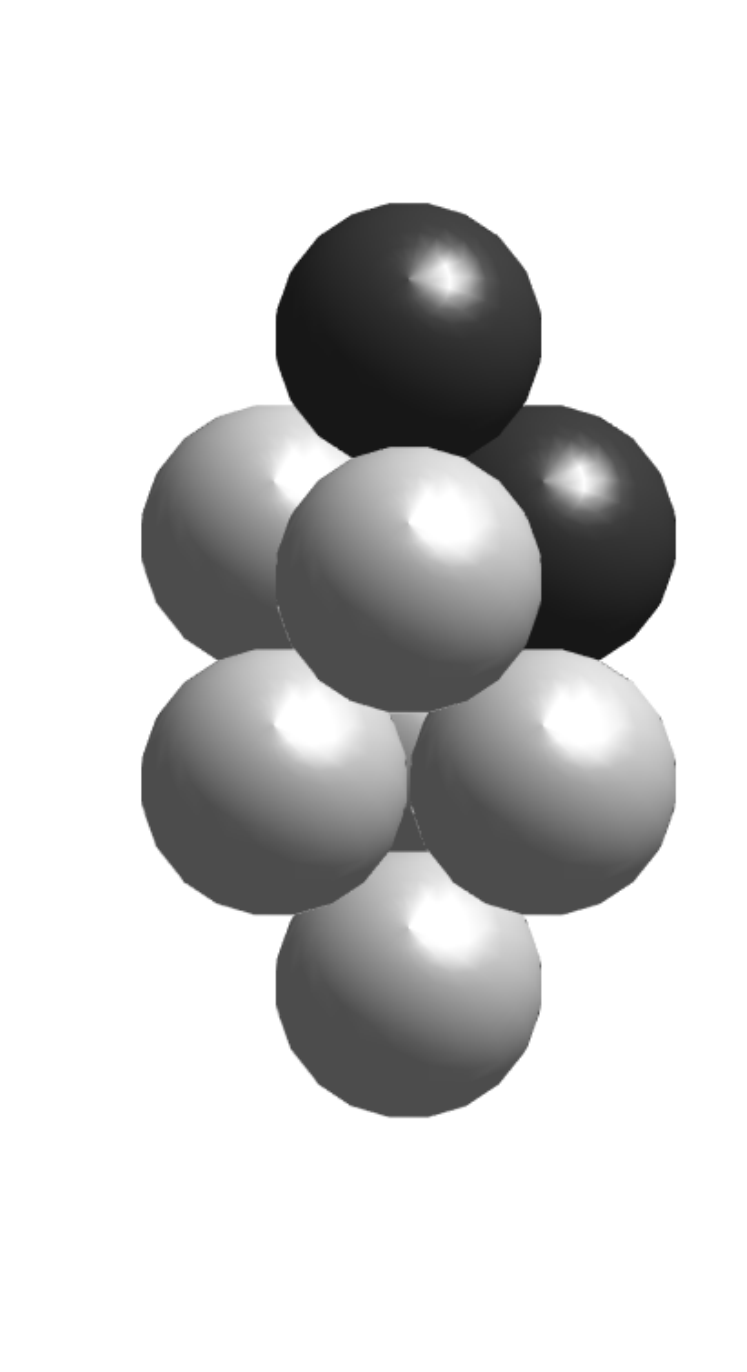}
\includegraphics[trim=70mm 70mm 70mm 70mm, clip, width=0.19\textwidth]{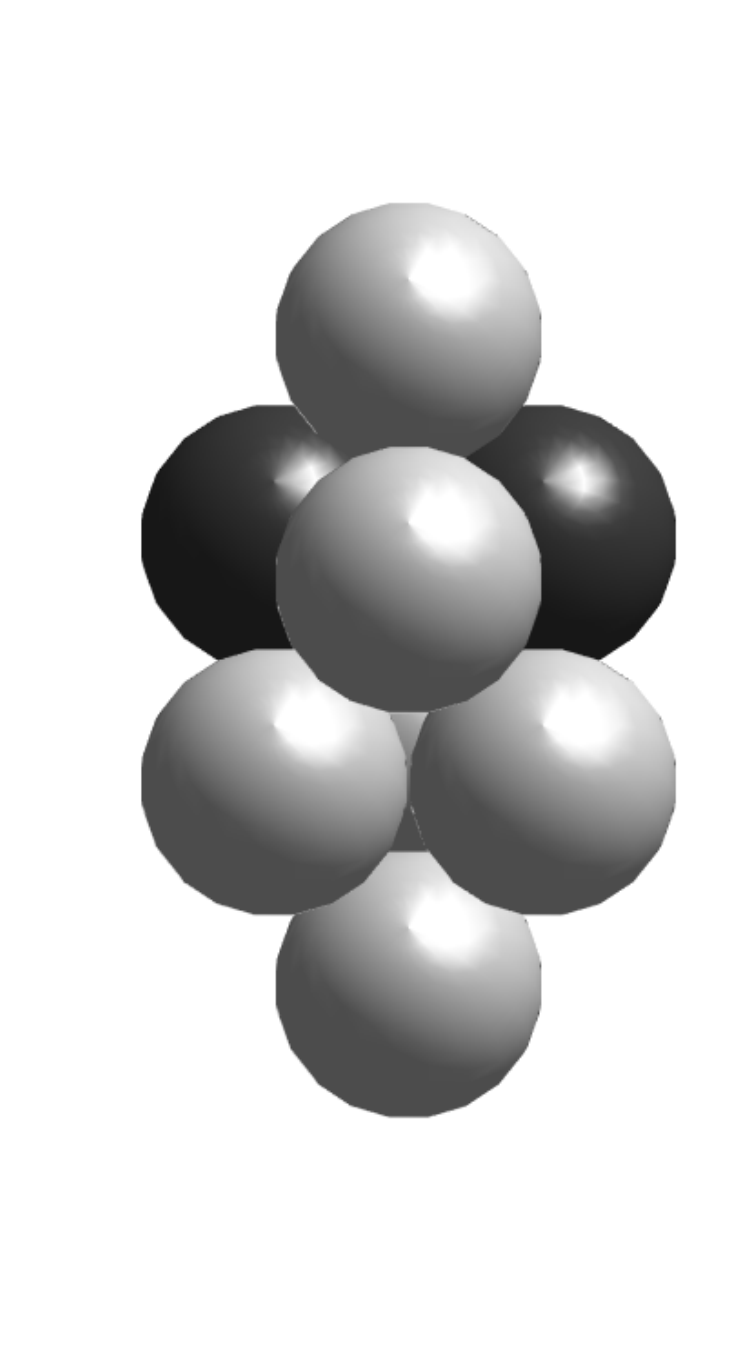}
\includegraphics[trim=70mm 70mm 70mm 70mm, clip, width=0.19\textwidth]{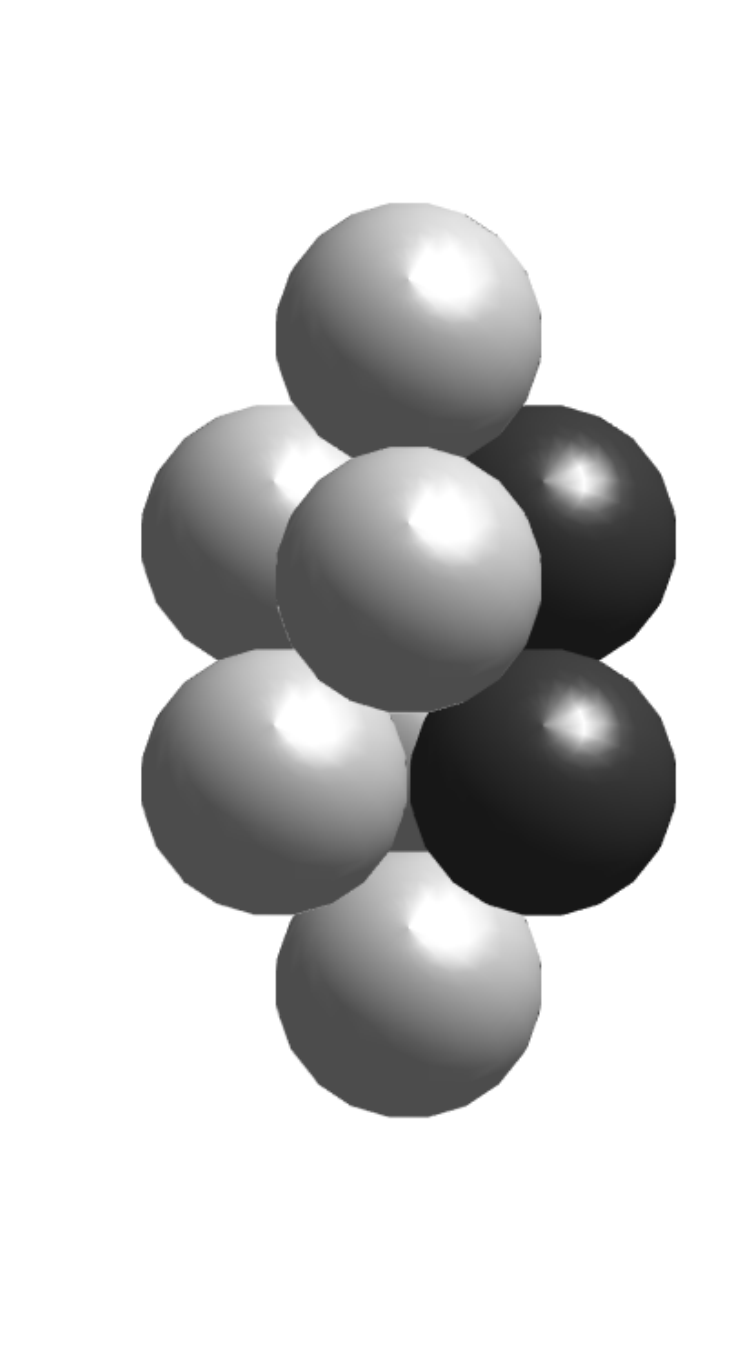} \\
\raggedright \hspace{2mm} (f) \hspace{18mm} (g) \hspace{18mm} (h) \hspace{18mm} (i) \hspace{18mm} (j)\\  \includegraphics[trim=70mm 70mm 70mm 70mm, clip, width=0.19\textwidth]{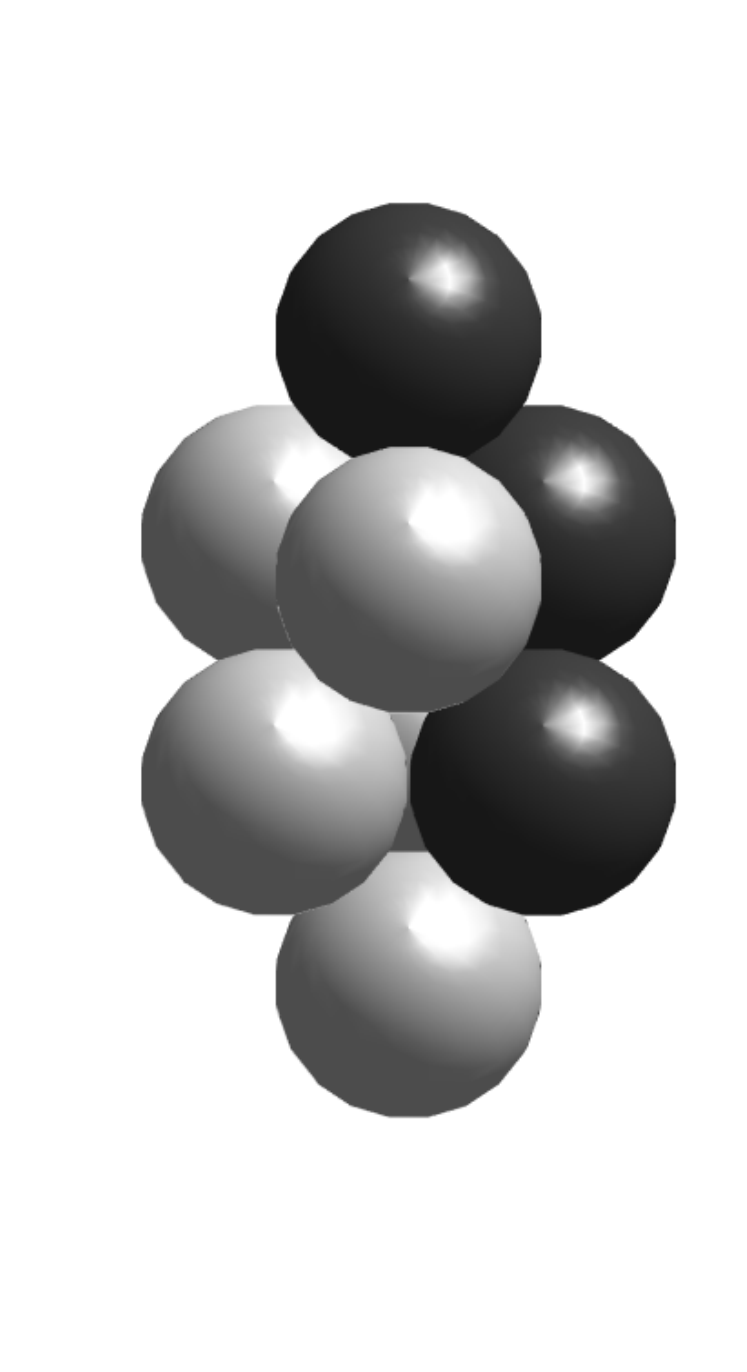}
\includegraphics[trim=70mm 70mm 70mm 70mm, clip, width=0.19\textwidth]{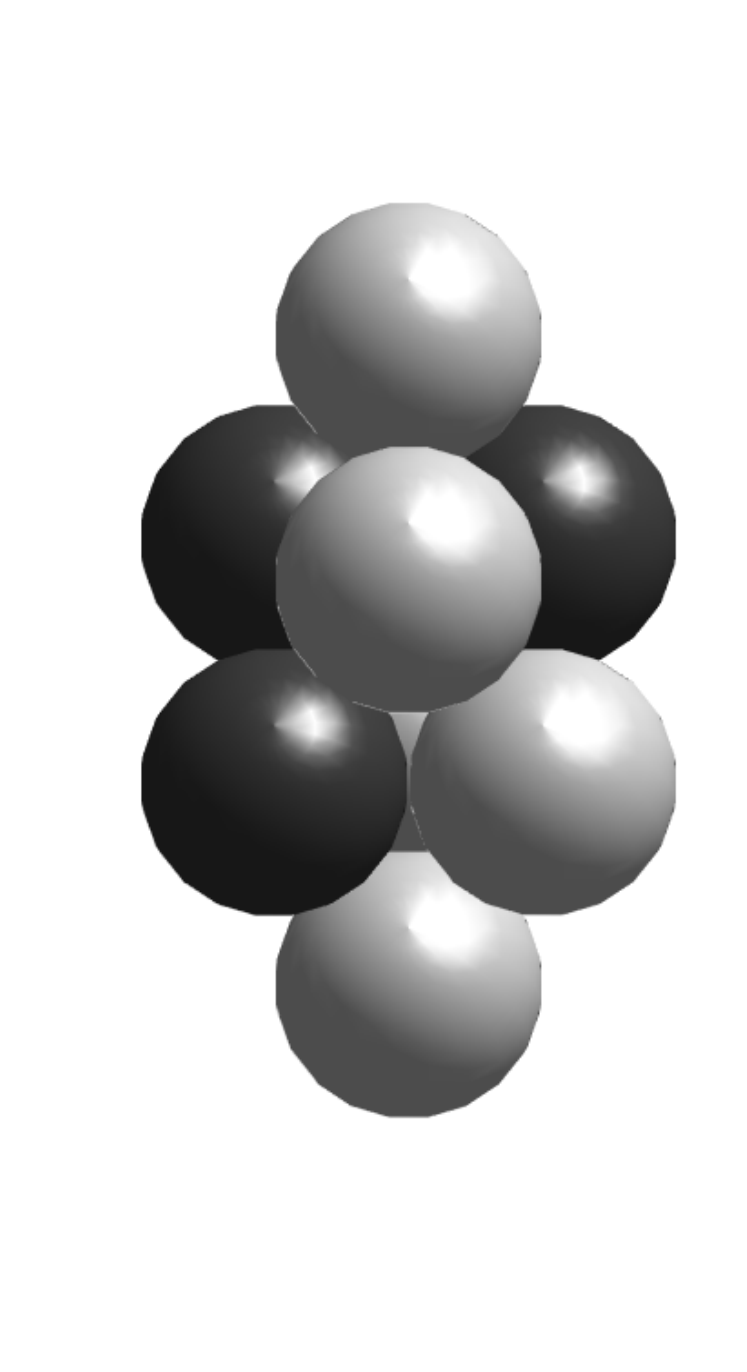}
\includegraphics[trim=70mm 70mm 70mm 70mm, clip, width=0.19\textwidth]{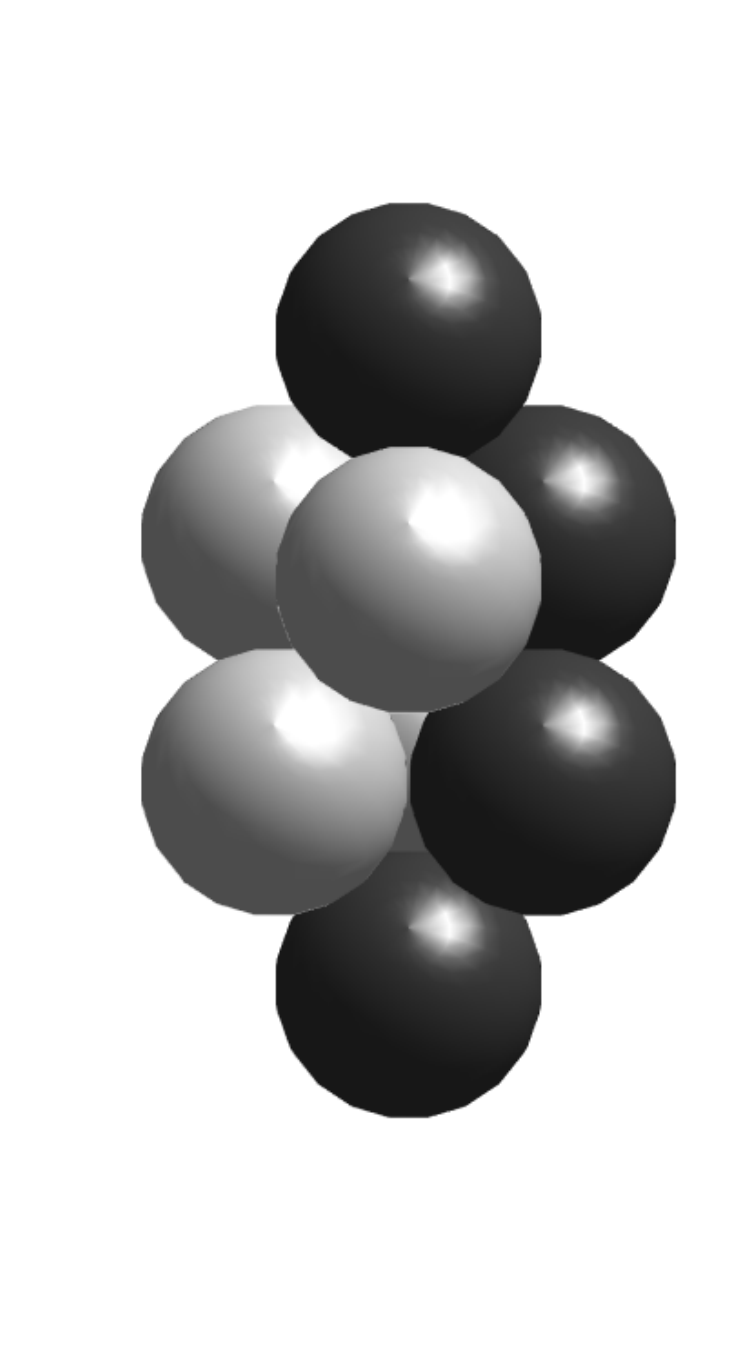}
\includegraphics[trim=70mm 70mm 70mm 70mm, clip, width=0.19\textwidth]{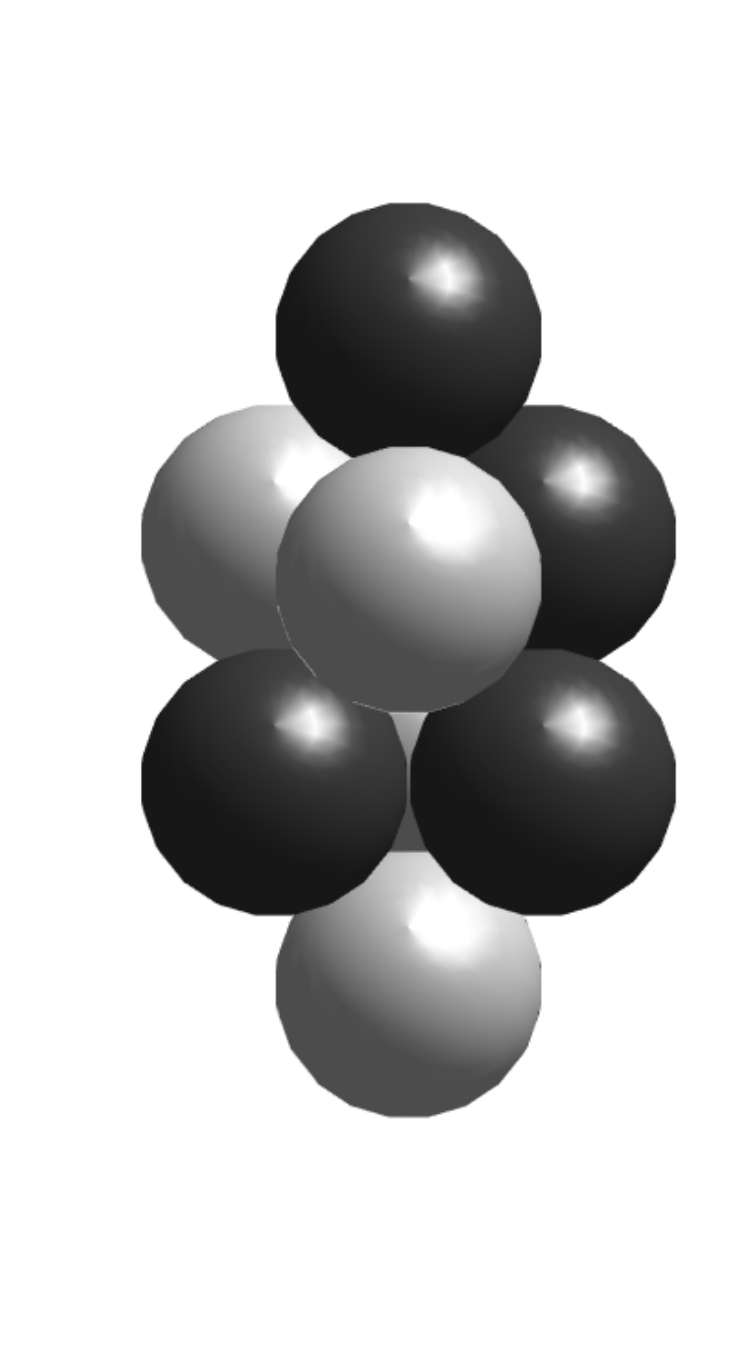}
\includegraphics[trim=70mm 70mm 70mm 70mm, clip, width=0.19\textwidth]{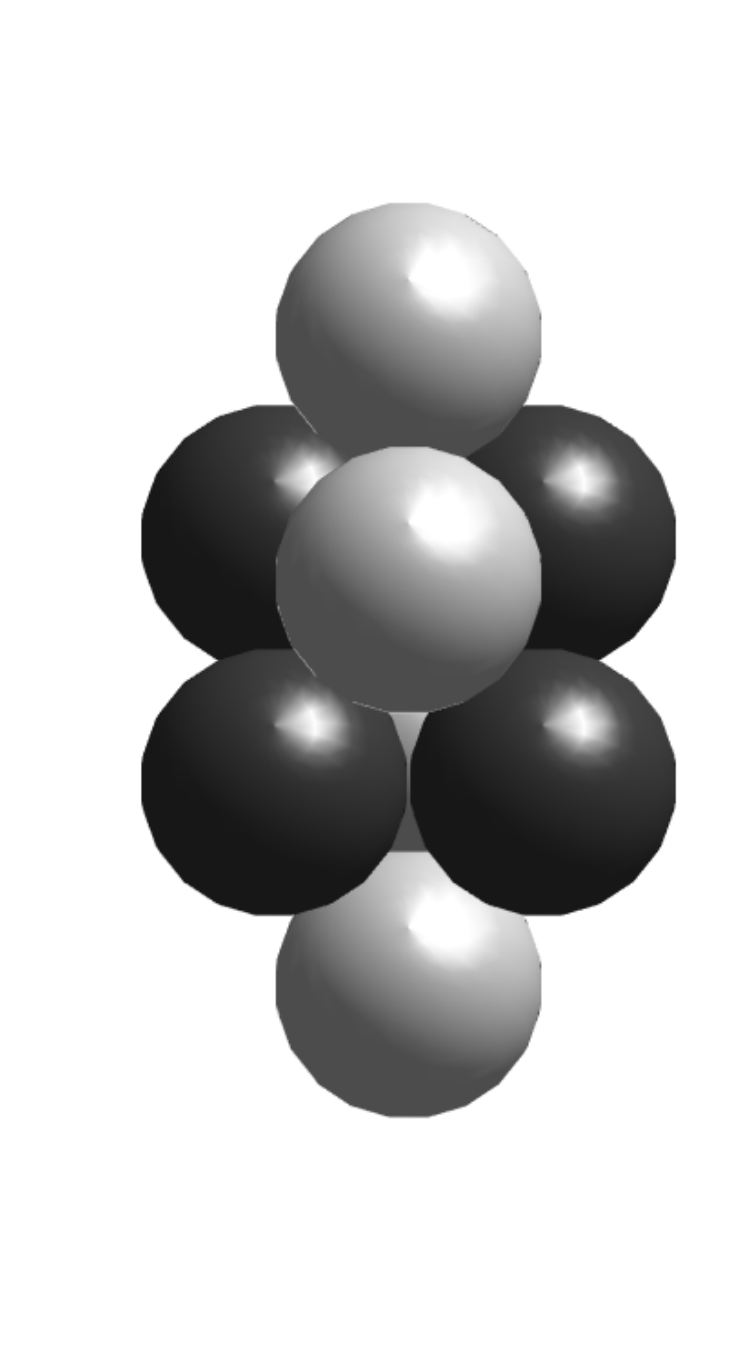}
\caption{The ten irreducible combinations of missing core sites, with missing: (a) end (E); (b) body (B); (c) end-body (EB); (d) body-body same level (BBs); (e) body-body different level (BBd); (f) end-body-body (EBB); (g) body-body-body (BBB); (h) end-body-body-end (EBBE); (i) end-body-body-body (EBBB); and (j) body-body-body-body (BBBB) The probability of stability of these ten configurations can be used to build the probability of stability for all other configurations.}
\label{fig:combos3D}
\end{figure}

To simplify the problem, different combinations of unoccupied sites which do not share any neighbors can be factorized. For example, if both the top and bottom particles are missing, each with nine neighboring sites, this yields $2^{18}$ configurations to check. However, since the top and bottom sites do not share neighbors, the probabilities can be factorized ($2^9$ configurations each), and the probability of stability for both missing sites is equal to the square of the probability for one to be stable. There are ten irreducible configurations, shown in Fig.~\ref{fig:combos3D}, with missing end (E), body (B), end-body (EB), body-body same level (BBs), body-body different level (BBd), body-body-body (BBB), end-body-body (EBB), end-body-body-end (EBBE), end-body-body-body (EBBB), and body-body-body-body (BBBB). Once these combinations are known, all possible configurations can be constructed from them. 

The stability criteria for these ten configurations is then solved numerically by iterating through all possible neighbor configurations. This yields polynomials $f_E(p)$, $f_B(p)$, $f_{EB}(p)$, etc., for each different configuration from Fig. \ref{fig:combos3D}, which gives the probability that the neighbors of the unoccupied sites are occupied such that the whole configuration (all 44 sites) is stable (i.e. no missing triads). For example, the polynomial for one missing body site (with seven neighbors) is written:
\begin{equation}
f_B(p)=p^7+7p^6(1-p)+11p^5(1-p)^2+3p^4(1-p)^3.
\label{eqn:fB}
\end{equation}
This means that, for a single unoccupied ``body'' site, there is one stable configuration with (all) seven neighbors occupied, seven stable configurations with six occupied neighbors, eleven stable configurations with five occupied neighbors, and three stable configurations with four occupied neighbors. These can be more compactly represented by
\begin{equation}
{}^N(a,b)=b p^{a} (1-p)^{N-a},
\label{eqn:fnotation}
\end{equation} 
where $a$ is the number of occupied neighbors, $b$ the number of stable configurations, and $N$ as the total number of neighbors under consideration. With this notation, one can write:
\begin{eqnarray*}
f_{\rm B}&=& {}^7(7, 1)+ {}^7(6, 7)+ {}^7(5, 11) + {}^7(4, 3)\\
f_{\rm E}&=& {}^9(9, 1)+ {}^9(8, 9)+ {}^9(7, 21)+ {}^9(6, 11)\\
f_{\rm EB}&=& {}^{14}(14, 1)+ {}^{14}(13, 12)+ {}^{14}(12, 51)+ {}^{14}(11, 94)+ {}^{14}(10, 77)\\ &&+\,   {}^{14}(9, 26)+ {}^{14}(8, 3)\\
f_{\rm BBs}&=& {}^{13}(13, 1)+ {}^{13}(12, 12)+ {}^{13}(11, 50)+ {}^{13}(10, 86)+ {}^{13}(9, 61)\\ &&+\,   {}^{13}(8, 14)+ {}^{13}(7, 1)\\
f_{\rm BBd}&=& {}^{12}(12, 1)+ {}^{12}(11, 10)+ {}^{12}(10, 33)+ {}^{12}(9, 40)+ {}^{12}(8, 16)\\
f_{\rm EBB}&=& {}^{19}(19, 1)+ {}^{19}(18, 15)+ {}^{19}(17, 87)+ {}^{19}(16, 248) + {}^{19}(15, 367) \\ &&+\, {}^{19}(14, 275)+ {}^{19}(13,95)+{}^{19}(12, 12)\\
f_{\rm BBB}&=& {}^{18}(18, 1)+ {}^{18}(17, 15)+ {}^{18}(16, 87)+ {}^{18}(15, 246)+ {}^{18}(14, 355) \\ &&+\, {}^{18}(13, 256) + {}^{18}(12, 82)+ {}^{18}(11, 8)\\
f_{\rm EBBE}&=& {}^{26}(26, 1)+ {}^{26}(25, 20)+ {}^{26}(24, 166)+ {}^{26}(23, 746)+ {}^{26}(22, 1989) \\ &&+\,  {}^{26}(21, 3244) + {}^{26}(20,3229)+ {}^{26}(19, 1918)+ {}^{26}(18, 658) \\ &&+\, {}^{26}(17, 120) + {}^{26}(16, 9)\\
f_{\rm EBBB}&=& {}^{25}(25, 1)+ {}^{25}(24, 20)+ {}^{25}(23, 166)+ {}^{25}(22, 744)+ {}^{25}(21, 1967) \\ &&+\,  {}^{25}(20, 3157) + {}^{25}(19,3067)+ {}^{25}(18, 1759)+ {}^{25}(17, 569) \\ &&+\, {}^{25}(16, 94) + {}^{25}(15, 6)\\
f_{\rm BBBB}&=& {}^{22}(22, 1)+ {}^{22}(21, 16)+ {}^{22}(20, 104)+ {}^{22}(19, 352)+ {}^{22}(18, 664) \\ &&+\,  {}^{22}(17, 704) + {}^{22}(16,416)+ {}^{22}(15, 128)+ {}^{22}(14, 16)\\
\end{eqnarray*}

Finally, the full probability of mechanical stability for the eight-particle cluster can now be constructed with consideration for all 36 neighbors. To do this, I organize by the number of occupied core sites, $m$, which gives a prefactor of $p^m (1-p)^{8-m}$ (i.e. the probability of $m$ occupied sites). Then, all possible configurations can be constructed by combining the ten irreducible configurations. For example, the $m=7$ term should account for two ways to have an end-particle missing ($2f_E$), plus six ways to have a body-particle missing ($6f_B$). The full RG transformation, $p'=R_{b=2}^{\rm (3D)}(p)$ is written:
\begin{eqnarray}
p'&=& R_{b=2}^{\rm (3D)}(p)\nonumber \\
&=& p^8+p^7(1-p)[2f_{\rm E}+6f_{\rm B}]\nonumber \\
&& +\,p^6(1-p)^2[(f_{\rm E})^2+6f_{\rm E} f_{\rm B}+6f_{\rm EB}+3(f_{\rm B})^2+6f_{\rm BBd}+6f_{\rm BBs}]\nonumber \\
&& +\, p^5(1-p)^3[6f_{\rm EB}f_{\rm E}+12f_{\rm EBB}+6f_{\rm EB}f_{\rm B}+6f_{\rm BBs}f_{\rm E}+12f_{\rm BBB}]\nonumber \\
&& +\, p^4(1-p)^4[3(f_{\rm EB})^2+6f_{\rm EBBE}+12f_{\rm EBBB}+3f_{\rm BBBB}]
\end{eqnarray}
\begin{figure}
    \includegraphics[scale=0.5]{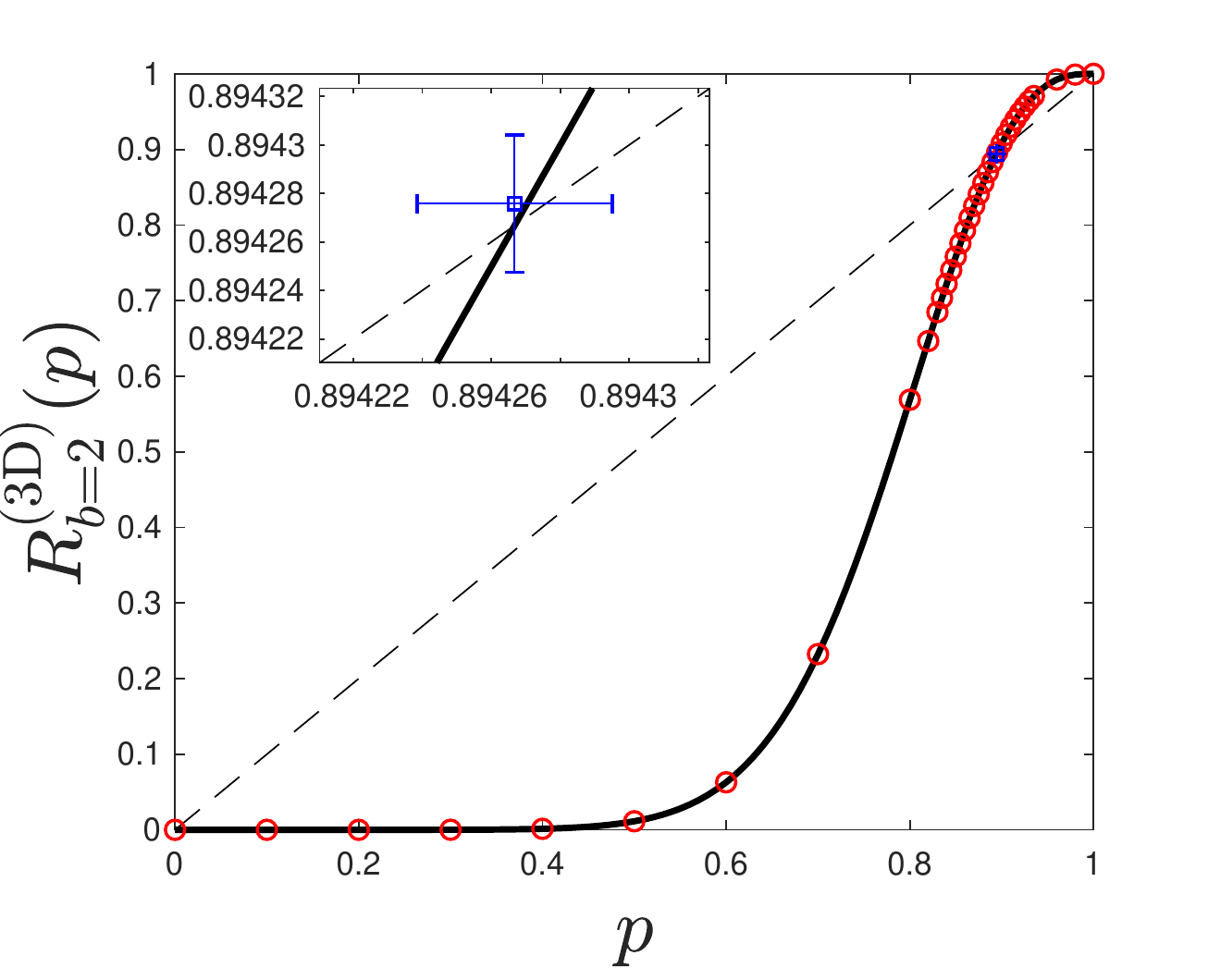}
    \caption{The function $R_{b=2}^{\rm (3D)}$ is plotted with the solid black line, and the dashed line shows a line of slope 1. The red circles indicate the average of $N$ Monte Carlo simulations, as described in the text, and the blue square shows the result of Monte Carlo simulations at the critical point. The inset shows a closeup near $p=p_c$. Error on the mean is proportional to $1/\sqrt{N}$, so the horizontal error bars have magnitude $p/\sqrt{N}$, and the vertical error bars have magnitude $p'/\sqrt{N}$, where $N=10^9$.}
    \label{fig:R_b_3D}
\end{figure}
This polynomial is plotted in Fig.~\ref{fig:R_b_3D} along with a dashed line representing $p'=p$. As in 2D, these two lines cross at $p=0$ and $p=1$ (the trivial fixed points) and at the critical point, where $R_{b=2}^{\rm (3D)}(p_c)=p_c$. Solving numerically yields
\begin{equation}
p_c=0.894267...
\label{eqn:pc3D}
\end{equation}
As in two dimensions, multiplying by the packing density of an FCC lattice of spheres, $\phi_L=\pi\sqrt{2}/6$, yields the critical packing fraction:
\begin{equation}
\phi_{c,3D} =p_c\phi_L =0.6622...
\label{eqn:phiJ3D}
\end{equation}
This value is similar to the random close packing value for monodisperse frictionless spheres in three dimensions, $\phi_{J,3D}\approx 0.64$. 

For the divergence of the correlation length, equation \ref{eqn:nu2} yields a value for the critical exponent, $\nu$, in three dimensions. Again, note that $b=2$ since there are $8=2^3$ particles.
\begin{equation}
\nu = \frac{\log b}{\log \frac{dR_b(p)}{dp}\vert _{p=p_c}}=\frac{\log 2}{\log 2.5055...}=0.7547...
\label{eqn:nu2D}
\end{equation}
This value is similar to O'Hern, et al.~\cite{ohern2003}, $\nu \approx 0.7$, but will likely depend on the choice of lattice and unit cell, as in 2D.

The above calculation is very complex. Thus, Monte Carlo simulations were performed to confirm the shape of the polynomial $R_{b=2}^{\rm (3D)}(p)$ in 3D as well as the critical values $p_c$ and $\nu$. In these simulations, the 44 sites that comprise the 8-site core plus 36 neighbors were filled randomly, where the probability that any individual site would be filled was equal to $p$. A configuration was denoted stable according to the criteria used to derive $R_{b=2}^{\rm (3D)}(p)$ above. Specifically, a configuration is stable if there are no sets of three neighboring unoccupied sites that both (1) form an equilateral triangle and (2) have one of the three unoccupied sites in the 8-site core (that is, a missing equilateral triangle comprised totally from neighbors is still considered stable). This process was repeated $N$ times for each $p$ at many values of $p$, and $R_{b=2}^{\rm (3D)}(p)$ was estimated as the number of stable configurations divided by $N$ at that $p$. For most values of $p$, $N=10^7$, but $N=10^9$ was used at selected values (including the critical point) to ensure convergence. The results of these simulations are shown as red circles in Fig.~\ref{fig:R_b_3D}. A blue square denotes the same Monte Carlo procedure performed at the critical point in Eq.~(\ref{eqn:pc3D}). This returns the same value, $p_c = R_{b=2}^{\rm (3D)} (p_c)$ within the expected error, confirming that the calculation shown above is correct.

Due to the complexity of this calculation, other choices for unit cells or lattices were not considered. However, as I show above in the 2D solution, making the unit cell larger (smaller) causes $\phi_c$ and $\nu$ to increase (decrease) somewhat. This is due to the fact that the stability criterion for larger clusters is more difficult to achieve. The 3D solution should exhibit the same general behavior. Regarding other lattices, the HCP lattice is very similar to the FCC lattice (with the same $\phi_{\rm xtal}$), so a similar solution may apply there. The BCC lattice, for example, would not make a good candidate for this approach, since a single missing lattice site in a BCC lattice is unstable. The BCC lattice also has $\phi_{\rm xtal} \approx 0.52$, which below the known packing fraction of disordered sphere packings. Additionally, it is inherently unstable if layers are allowed to slip in a transverse manner. 

\section*{Discussion} \label{sec:discussion}

Here, I have shown how a real-space renormalization group might be developed for the unjamming of a close-packed lattice of spheres with some percentage of the spheres randomly removed. For the solution I present here to be explicitly and fully connected to jamming in disordered systems, two questions must be answered. First: does the RVL state relate to jamming at all? For example, is not immediately obvious whether the RVL systems considered here will, even if they are unstable, relax to the same disordered states as in typical jamming studies. It is likely that the RVL model in 3D would ``relax" (by whatever protocol is used) to a disordered state, while a 2D system of monodisperse disks will more likely form crystalline states~\cite{reis2006crystallization}. This is because the hexagonally packed lattice is both the locally and globally preferred packing in 2D. In 3D, this is not the case, since the locally preferred packing is an icosohedron with five-fold symmetry, which cannot form a lattice~\cite{nelson1989polytetrahedral}. However, recent work has shown that disordered 2D packings of monodisperse disks are possible~\cite{atkinson2014existence}. Preliminary discrete-element simulations were performed (not shown here) of 3D RVL systems using 1000 sites ($10 \times 10 \times 10$) with soft spheres in a fully periodic cubic cell (the actual number of spheres is less than 1000, since some are randomly removed). When these systems are prepared in an RVL state and then subjected to athermal, quasistatic compression with fully overdamped dynamics, similar to~\cite{ohern2003}, simulations show that the RVL systems with $\phi < 0.64$ do relax and compress to form disordered packings at $\phi \approx 0.64$, while RVL systems with $\phi>0.64$ do not relax (i.e., they remain in their initial configuration but with additional compression energy). This gives some confidence that the RVL state does relate to jamming, at least in the sense that it may be a valid initial condition from which to prepare jammed states. A complete characterization of these states, including system-size dependence and a comparison to states generated via other jamming protocols, will be the subject of a future study.

The second question is: if we assume that the RVL model does relate to jamming of disordered systems in the manner described above, does the RG transformation faithfully describe the way that the system relaxes from an RVL state to a disordered, jammed state? For the RG I present here, the answer to this question is obviously no, at least not exactly. The primary reason is that the RG transformations I present here suffer from the same flaw as Kadanoff's initial block-spin approach, in that they assume that the interactions among single sites (before coarse graining) are the same as the interactions among super-sites (after coarse graining). This is obviously false since, for example, the interaction of a 3D super-site on the FCC lattice and its twelve super-site neighbors is much more complex than a single sphere and its twelve neighboring sites. Additionally, the honeycomb lattice is collectively unstable~\cite{torquato2001multiplicity}, but the RG presented here defines it as completely stable. However, the RVL system assumes that the voids are random, while the voids in the honeycomb lattice are not random. Thus, future work could certainly improve upon the block-spin-like RG transformation I present here to find a more accurate way treat interactions at larger length scales. A complete RG approach should also capture other quantities that discontinuously appear or vanish at jamming, such as the bulk and shear moduli~\cite{goodrich2016scaling}. These calculations are outside the scope of this paper, and it is not clear how an RG approach to the RVL model would treat these quantities.


\section*{Acknowledgements}
I would like to acknowledge contributions from Bob Behringer, Richard Palmer, and Josh Socolar, who each helped me develop this idea through many informal discussions while I was a graduate student at Duke.

\section*{Compliance with Ethical Standards}
Conflict of Interest: The author declares that he has no conflict of interest. This research involved no human participants or animals.


\raggedright
\bibliographystyle{spphys}       
\bibliography{references.bib}   

\end{document}